\documentstyle[12pt,epsfig]{article}
\def\be {\begin{equation}}
\def\ee {\end{equation}}
\def\bea {\begin{eqnarray}}
\def\eea {\end{eqnarray}}
\def\nn {\nonumber}

\newcommand {\om}{{\omega}}

\begin{document}

\thispagestyle{empty}
\vskip 15pt

\begin{center}
{\Large {\bf Radiative energy loss in an anisotropic {\em Quark-Gluon-Plasma}}}
\renewcommand{\thefootnote}{\alph{footnote}}

\hspace*{\fill}

\hspace*{\fill}

{ \tt{
Pradip Roy\footnote{E-mail address:
pradipk.roy@saha.ac.in}
and
Abhee K. Dutt-Mazumder\footnote{E-mail address:
abhee.dm@saha.ac.in}
}}\\

\small {\em Saha Institute of Nuclear Physics \\
1/AF Bidhannagar, Kolkata - 700064, INDIA}
\\

\vskip 40pt

{\bf ABSTRACT}
\end{center}

\vskip 0.5cm

We calculate radiative energy loss of heavy and light quarks
in anisotropic medium (static) in first order opacity expansion.
Such an anisotropy can result from the 
initial rapid longitudinal expansion of the matter, created in relativistic 
heavy ion collisions. 
Significant dependency of the energy loss 
on the anisotropy parameter ($\xi$) and the direction of propagation
of the partons with respect to the anisotropy axis is found. 
It is shown that the 
introduction of early time momentum-space anisotropy can 
enhance the fractional energy loss in the direction of the anisotropy, whereas
it decreases when the parton propagates perpendicular to the direction
of the anisotropy. 
 
\vskip 30pt

\section{Introduction}

One of the goals for the ongoing relativistic heavy ion 
collision experiments at the Relativistic Heavy Ion 
Collider (RHIC) and the upcoming experiments at CERN Large Hadron 
Collider (LHC) is to produce quark gluon plasma (QGP) and study its properties.
According to the prediction 
of lattice quantum chromodynamics, QGP is expected to 
be formed when the temperature of nuclear matter is raised above 
its critical value, $T_c\sim 170$ MeV, or equivalently the energy 
density of nuclear matter is raised above $1~GeV/fm^{3}$
~\cite{PRC72_ref1}. The possibility of QGP formation at RHIC 
experiment, with initial density of $5~GeV/fm^{3}$ initial density  
is supported by the observation of high $p_T$ hadron suppression
(jet-quenching) in the central Au-Au collisions compared to the 
binary-scaled hadron-hadron collisions~\cite{jetquen}. 
Apart form jet-quenching, 
several possible probes have been studied in order to characterize the 
properties of QGP.

However, many properties of QGP are still poorly understood. The most 
debated question is whether the matter formed in the relativistic heavy ion
collisions is in thermal equilibrium or not. The measurement of elliptic
flow parameter and its theoretical explanation suggest that the matter
quickly comes into thermal equilibrium 
(with $\tau_{\rm therm} < $ $1$ fm/c, where $\tau_{therm}$ is the time of 
thermalization)~\cite{PRC75_ref1}.
On the contrary, perturbative estimation suggests relatively
slower thermalization of
QGP~\cite{PRC75_ref2}. However, recent hydrodynamical 
studies~\cite{0805.4552_ref4} have shown that 
due to the poor knowledge of the initial conditions there is a
sizable amount of uncertainty in the estimate of thermalization or 
isotropization time. It is suggested that (momentum) anisotropy driven
plasma instabilities may speed up the process of 
isotropization~\cite{plb1181993}, in that case one is allowed to use 
hydrodynamics for the evolution of the matter. However, instability-driven 
isotropization is not yet proved at RHIC and LHC energies. 

In absence of a theoretical proof favoring the rapid
thermalization and the uncertainties in the hydrodynamical fits of
experimental data, it is very hard to assume hydrodynamical
behavior of the system from the very beginning. Therefore, 
it has been suggested to look for some observables which are sensitive to the
early time after the collision. For example, jet-quenching vis-a-vis energy
loss of partons could be an observable where the initial state
momentum anisotropy can play important role. This is the
issue that we address here. 

It is known that the energy loss of partons (also dubbed as
'jet-quenching') in QCD plasma can proceed in two ways:
by two-body scattering and also via gluon radiation. These
are known as collisional and radiative energy loss respectively.
The phenomena of jet-quenching has been investigated
by various authors~\cite{jetquen}
More recently, the non-photonic single electron data shows more suppression
than expected which cannot be explained by radiative loss alone.
A substantial
amount of work has been done to look into this issue in recent
times~\cite{jetquen}.

It is to be noted that the existing calculations on energy loss
have been performed in isotropic QGP which is true immediately
after its formation~\cite{5of4552}. However, subsequent rapid
expansion of the matter along the beam direction causes faster cooling in
the longitudinal direction than in the transverse
direction~\cite{PRC75_ref2}. As a result, the system becomes anisotropic
with $\langle{p_L}^2\rangle << \langle{p_T}^2\rangle$ in the local rest frame.
At some later time when the effect of parton interaction rate
overcomes the plasma expansion rate, the system returns to the
isotropic state again and remains isotropic for the rest of the period.
Thus, during the early stage the plasma remains anisotropic and any calculation
of energy loss should, in principle, include this aspect.
The collisional energy loss in anisotropic media for heavy fermions has been 
calculated in Refs.~\cite{strick1,strick}. 
In these calculations it is found that the deviations from the isotropic
results are of the order of 10\% for $\xi=1$ and of the order of
20\% for $\xi=10$. It is observed that the collisional energy loss
varies with the angle of propagation by upto 50\%.

Recently in \cite{baier}, the transport coefficient ${\hat q}$ has been 
calculated in anisotropic media, which in turn, affects the radiative
energy loss. 
We here attempt to provide quantitative
estimate of radiative energy loss by modifying the static scatterer 
model~\cite{magdalenaprc} appropriate for anistropic media.


The other interesting aspect which, in recent years, has attracted considerable
attention is the possibility of the growth of unstable modes in anisotropic
plasma~\cite{mrow1}. For example, in \cite{abhijit} the authors calculate
${\hat q}$ for a two-stream plasma and show that the momentum broadening
grows exponentially in time  
as the spontaneously growing fields exert an exponentially growing
influence on the propagating parton. This momentum broadening
of a fast parton which radiates gluons due to the scattering off 
the plasma constituents therefore controls the radiative energy 
loss~\cite{baier1}. In an evolving plasma this is
an important component, which, however, is not included in the 
present manuscript. Therefore the results we report here can be considered
to be something like a zeroth order approximation.

The plan of the paper is the following. In section 2 we briefly
mention how to calculate the 2-body potential in an
anisotropic media along with the  modified expression
for the fractional energy loss. Section 3 will be devoted
to discuss the results. Finally, we conclude in section 4.

\section{Formalism}   

\begin{figure}[h]
\begin{center}
\epsfig{file=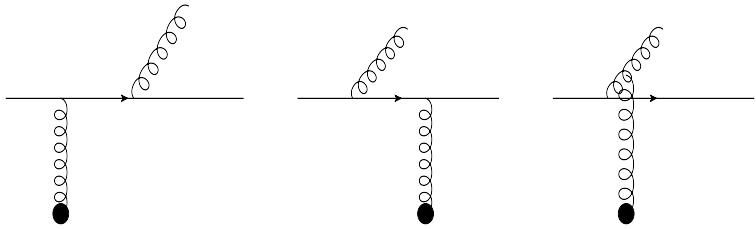,width=10cm,height=3cm,angle=0}
\end{center}
\caption{Feynman diagrams contributing to the soft gluon radiation in
a static medium to first order in opacity}
\label{fig1a}
\end{figure}

In this section, we recapitulate the basic formalism of the radiative
energy loss of a fast moving parton in an infinitely extended static isotropic 
plasma~\cite{mplb,mnpa,magdalenaprc}. As in Ref.~\cite{magdalenaprc}
we restrict ourselves to the radiative energy loss quarks at first
order in opacity involving three diagrams as shown in Fig.(\ref{fig1a})
where
we assume that an on-shell heavy quark produced in the remote past
is propagating through an infinite QCD medium that consists of randomly
distributed static scattering centers. In the original Gyulassy-Wang
formalism~\cite{gul} static interactions are modeled here as color-screened Yukawa
potential originally developed for the isotropic QCD medium given by
\bea
V_n&=&V(q_n)e^{iq_n.x_n}\nn\\  
   &=&2\pi \delta(q^0)v(q_n)e^{-iq_n.x_n}T_{a_n}(R)\otimes T_{a_n}.
\eea
with $v(q_n)=4\pi\alpha_s/(q_n^2+\mu^2)$, where $\mu$ is the Debye
mass. $x_n$ is the location of the $n$th scattering centre, $T$ (summed
over $a_n$) denotes the colour matrices of the parton and the scattering
centre. It is to be noted that the potential has been derived by using
hard thermal loop (HTL) propagator in QGP medium. In a plasma with momentum anisotropy the 
two body interaction, as expected, becomes direction dependent. It has 
been observed that, on distance scale of the order of the inverse Debye 
mass, the attraction for the quarks aligned along the direction of anisotropy 
is stronger than for transverse alignment \cite{dumitru08}. Therefore,
the radiative energy loss, will also depend on the direction of momentum
of the quarks emitting Bremstrahlung gluons. 
This necessitates the introduction of anisotropy dependent potential
to estimate the radiative energy loss in a plasma having anisotropic
momentum distribution.

The heavy-quark potential in an anisotropic plasma has recently been
calculated in \cite{dumitru08} for which one starts with retarded gluon
self-energy expressed as~\cite{blaizot} 
\be
\Pi^{\mu \nu}(P) = g^2\,\int\,\frac{d^3k}{(2\pi)^3}v^{\mu}\,
\frac{\partial f({\vec k})}{\partial K^\beta}
\left(g^{\nu \beta}-\frac{v^{\nu} P^\beta}{P\cdot v+i\epsilon}\right)
\label{se}
\ee
We have adopted the following notation for four vectors: 
$P^{\mu} = (p_0,{\vec p}) = (p_0,{\bf p},p_z)$, i. e. ${\vec p}$
(with an explicit vector superscript) describes a three-vector while
${\bf p}$ denotes the two-vector transverse to the $z$-direction.

To include the local anisotropy in the plasma, one has to 
calculate the gluon polarization tensor incorporating anisotropic distribution
functions of the medium. This subsequently can be used to construct
HTL corrected gluon propagator which, in general,
assumes very complicated from. Such an HTL propagator was first
derived in \cite{Romash} in time-axial gauge. Similar propagator
has also been constructed in \cite{dumitru08} to derive the heavy-quark
potential in an anisotropic plasma, which, as we know, is given by the 
Fourier transform of the propagator in the static limit.

The self-energy, apart from momentum $P^\mu$, also depends on a
fixed anisotropy vector $n^{\mu} (=(1,{\vec n}))$ and $\Pi^{\mu \nu}$
can be cast in a suitable tensorial basis appropriate
for anisotropic plasma in a co-variant gauge in the
following way~\cite{dumitru08}:
\be
\Pi^{\mu \nu} = \alpha\,A^{\mu \nu}+\beta\,B^{\mu \nu}+\gamma\,C^{\mu \nu}
+\delta\,D^{\mu \nu}
\ee
where the basis tensors are constructed out of $p^\mu$, $n^\mu$ and the
4-velocity of the heat bath $u^{\mu}$. The detailed expressions for
the quantities those appear in Eq.(3) can be found in Ref.~\cite{dumitru08}.
The anisotropicity enters through the distribution function,
\be
f(\vec p) = f_{\rm iso}(\sqrt{{\vec p}^2+\xi({\vec p}\cdot {\vec n})^2})
\ee 
where, the parameter $\xi$ is the degree of anisotropy 
parameter ($ -1 < \xi < \infty $) and  is given by
$\xi=\langle {\bf p^2} \langle/(2  \rangle p_z^2 \rangle)-1$. It is to be noted 
that $\xi$ can also be related to the shear viscosity~\cite{asakawa}.

Since the self-energy is symmetric and transverse, all the components
are not independent. After change of variables ($p^\prime = 
{\vec p}^2[1+\xi({\hat {\bf p}}
\cdot {\vec n})^2]$ the spatial components can be written as
\be
\Pi^{i j} = \mu^2\,\int\,\frac{d\Omega}{4\pi}\,v^i
\frac{v^l+\xi({\vec v}\cdot {\vec n})n^l}{1+\xi({\vec v}\cdot {\vec n})^2}
\left(\delta^{j l}+\frac{v^j p^l}{P\cdot v +i\epsilon}\right)
\ee
Now $\alpha, \beta, \gamma$ and $\delta$ are determined by the following
contractions:
\bea
p^i\,\Pi^{i j}\,p^j& =& {\vec p}^2\beta\nonumber\\
A^{i l}\,n^l\,\Pi^{i j}\,p^j& =& ({\vec p}^2-(n\cdot P)^2)\delta\nonumber\\
A^{i l}\,n^l\,\Pi^{i j}\,A^{j k}\,n^k& =& 
\frac{{\vec p}^2-(n\cdot P)^2}{{\vec p}^2}\alpha+\gamma)\nonumber\\
{\rm Tr}\Pi^{i j}& =& 2\alpha+\beta+\gamma
\eea
where the expressions for $\alpha, \beta, \gamma$ and $\delta$ are given
in Ref.~\cite{Romash}.

After knowing the gluon HTL self-energy in anisotropic
media the propagator can be calculated after some cumbersome
algebra~\cite{baier,dumitru08}: 
\bea
\Delta^{\mu\nu}&&=\frac{1}{(P^2-\alpha)}\big [A^{\mu\nu}-C^{\mu\nu} \big ]\nn\\
&& + \Delta_G \Big [(P^2-\alpha-\gamma)\frac{\om^4}{P^4}B^{\mu\nu}+
(\om^2-\beta) C^{\mu\nu}+\delta \frac{\om^2}{P^2}D^{\mu\nu} \Big] 
-\frac{\lambda}{P^4}P^\mu P^\nu
\eea
where
\bea
\Delta_G^{-1}=(P^2-\alpha-\gamma) (\om^2-\beta) -\delta^2 [P^2-(n\cdot P)^2]
\eea
Now the momentum space potential can be obtained from the static gluon
propagator in the following way,
\bea
v({\bf q},q_z,\xi)&=&g^2\,\Delta^{00}(\omega=0,{\bf q},q_z,\xi)
\nonumber\\
&=&g^2\,\frac{{\vec q}^2+m_\alpha^2+m_\gamma^2}
{({\vec q}^2+m_\alpha^2+m_\gamma^2)({\vec q}^2+m_\beta^2)-m_{\delta}^2}
\label{pot1}
\eea
where,
\bea
m_\alpha^2&=&-\frac{\mu^2}{2{\bf q}^2\sqrt{\xi}}\left[
q_z^2\tan^{-1}(\sqrt{\xi})-\frac{q_z{\vec q}^2}{\sqrt{{\vec q}^2+\xi{\bf q}^2}}
\tan^{-1}\left(\frac{\sqrt{\xi}q_z}{\sqrt{{\vec q}^2+\xi{\bf q}^2}}\right)\right]\nonumber\\
m_\beta^2&=& \mu^2\,\frac{
(\sqrt{\xi}+(1+\xi)\tan^{-1}(\sqrt{\xi}))({\vec q}^2+\xi{\bf q}^2)
+\frac{{\vec q}^2(1+\xi)}{\sqrt{{\vec q}^2+\xi {\bf q}^2}}
\tan^{-1}\left(\frac{\sqrt{\xi}q_z}{\sqrt{{\vec q}^2+\xi {\bf q}^2}}\right)}
{2\sqrt{\xi}(1+\xi)({\vec q}^2+\xi {\bf q}^2}
\nonumber\\
m_\gamma^2&=&-\frac{\mu^2}{2}\left[
\frac{
{\vec q}^2}{\xi {\bf q}^2+{\vec q}^2}
-\frac{1+\frac{2q_z^2}{{\bf q}^2}}{\sqrt{\xi}}\tan^{-1}(\sqrt{\xi})
+\frac{q_z{\vec q}^2(2{\vec q}^2+3\xi {\bf q}^2)}{\sqrt{\xi}(\xi{\bf q}^2
+{\vec q}^2)^{3/2}{\bf q}^2}\tan^{-1}\left(\frac{\sqrt{\xi}q_z}
{\sqrt{\vec q}^2+\xi{\bf q}^2}\right)
\right]\nonumber\\
m_\delta^2&=&-\frac{\pi \mu^2 \xi q_z {\bf q}{|\vec q|}}
{4(\xi{\vec q}^2+{\vec q}^2)^{3/2}}
\eea

with ${\vec q} = ({\bf q},q_z)$.
For general anisotropy vector ${\vec n}$ we have,
${\bf q} = {\vec q} - ({\vec q}\cdot {\vec n}){\vec n}$ and
${q_z} = {\vec q}\cdot {\vec n}$.

For, $q_z = 0$, the potential in anisotropic media simplifies to
\be
v({\bf q},\xi)=\frac{4 \pi \alpha_s}{{\bf q}^2+R(\xi)\mu^2} 
\ee
\label{pot2}

\be
R(\xi)=\frac{1}{2} \Big [\frac{1}{1+\xi} + \frac{\tan ^{-1}\sqrt{\xi}}{\sqrt{\xi}}\Big ]
\ee

For small anisotropicity and $q_z =  0$, the two body interaction can be written
as
\be
v({{\bf q},\xi<<1})=4 \pi\alpha_s \left[ \frac{1}{{\bf q}^2+\mu^2}+
\frac{2}{3}\frac{\mu^2\xi}{({\bf q}^2+\mu^2)^2}\right ]
\ee

Now in Fig.~(\ref{fig1a}) the parton scatters with one of the
colour centre with the momentum $Q = (0,{\bf q},q_z)$ and radiates a
gluon with momentum $K = (\omega,{\bf k},k_z)$. The method for calculating
the amplitudes of the diagrams depicted in Fig.~(\ref{fig1a})
is discussed in Refs.~\cite{mplb,mnpa} and we shall quote the main
results only.
The
quark energy loss is calculated by folding the rate of gluon radiation
($\Gamma (E)$) with the gluon energy by assuming $\omega+q_0 \approx
\omega $. In this approximation one finds,

\bea
\frac{dE}{dL}&=
&\frac{E}{D_R} \int x dx \frac{d\Gamma}{dx}
\eea
Here $D_R$ is defined as $[t_a,t_c][t_c,t_a]=C_2(G)C_RD_R$
where $C_2(G)=3$, $D_R=3$ and $[t_a,t_c]$ is a color commutator
(see ~\cite{magdalenaprc} for details). $x$ is the longitudinal
momentum fraction of the quark carried away by the emitted gluon. 
\vskip 0.5in
\begin{figure}[h]
\begin{center}
\epsfig{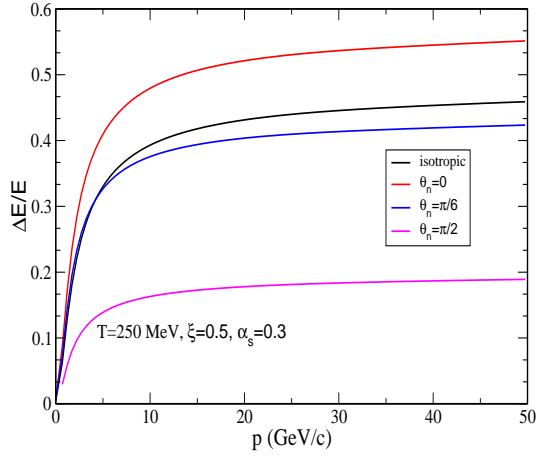}
\end{center}
\caption{Fractional energy loss for light quark when mean free is given
is given by Eq.(19).}
\label{fig1}
\end{figure}

where in anisotropic media we have,

\bea
x \frac{d\Gamma}{dx}&=&\frac{C_R\alpha_s}{\pi} \frac{L}{\lambda}
\int \frac{d^2{\bf k}}{\pi}\frac{d^2{\bf q}}{\pi} |v({\bf q},q_z,\xi)|^2 
\frac{\mu^2}{16\pi^2\alpha_s^2}
\Big [\frac{{\bf {k+q}}}{({\bf{k+q}})^2+\chi^2}-\frac{{\bf k}}{{\bf k}^2+\chi} \Big ]^2
\eea

In last expression, $v({\vec q},\xi)$ is the two body quark quark potential 
given by Eq.(9) and $\chi = m_q^2x^2+m_g^2$, where $m_g^2=\mu^2/2$ and
$m_q^2 = \mu^2/6$.

In the present case, we assume that the parton is propagating along the
$z$-direction and the anisotropy vector ${\vec n}$ makes an angle
$\theta_n$ with the $z$-axis, i.e. ${\vec n} = 
(\sin\theta_n, 0,\cos\theta_n)$. Thus $\theta_n$ describes the direction
of propagation of the parton with respect to the anisotropy axis. In such
case we replace ${\bf q}$ and $q_z$ in Eq.(9) by
${\bf q} \rightarrow \sqrt{{\bf q}^2-{\bf q}^2\sin^2\theta_n \cos^2\phi}$
and $q_z \rightarrow |{\bf q}|\cos\phi \sin\theta_n$, where
${\bf q} = (|{\bf q}|\cos\phi,|{\bf q}|\sin\phi)$.  

For arbitrary $\xi$ the radiative energy loss can be written as
\bea
\frac{\Delta E}{E}&=&\frac{C_R\alpha_s}{\pi^2} \frac{L\mu^2}{\lambda}
\int dx d^2{\bf q} \frac{|v({\bf q},q_z,\xi)|^2}{16\pi^2\alpha_s^2}
\Big [   
-\frac{1}{2}\nn\\
&&-\frac{k_m^2}{k_m^2+\chi} + 
\frac{{\bf q}^2-k_m^2+\chi}{2 \sqrt{{\bf q}^4+ 2 {\bf q}^2 (\chi-k_m^2)}+(k_m^2+\chi)^2
}+\nn\\
&&
\frac{{\bf q}^2+ 2 \chi} {{\bf q}^2 \sqrt{1+ \frac{4\chi}{{\bf q}^2} }}\ln
\Big ( 
\frac{k_m^2+\chi}{\chi} 
\frac{({\bf q}^2+3 \chi) + \sqrt{ 1+ 
\frac{4\chi}{{\bf q}^2} } ({\bf q}^2+ \chi)}{({\bf q}^2-k_m^2+3 \chi) 
+  \sqrt{1+ \frac{4\chi}{{\bf q}^2} } \sqrt{{\bf q}^4+ 2 {\bf q}^2 (\chi-k_m^2)}+(k_m^2+\chi)^2 }
\Big)
\Big ]\nn\\
\eea
In the above expression, $\lambda$ denotes the average mean free path
of the quark given by
\begin{equation}
\frac{1}{\lambda} = \frac{1}{\lambda_g}+ \frac{1}{\lambda_q}
\end{equation}
which in this case would be $\xi$ dependent. In the last expression
$\lambda_g$ and $\lambda_q$ correspond to the contributions coming from
$q$-$g$ and $q$-$q$ scatterings.

Explicitly with
Eq.(11) we have,
\begin{equation}
\lambda_i^{-1} = \frac{C_R C_2(i) \rho_i}{d_A}\,
\int\,d^2{\bf q}\,\frac{4\alpha_s^2}{({\bf q}^2+R(\xi)\mu^2)^2}.
\end{equation}
where $C_R=4/3$, $C_2(i)$ is the cashimir for $d_i$-dimensional
representation and $C_2(i)=(N_c^2-1)/(2N_c)$ for quark and
$C_2(i)=N_c$ for gluon scatterers. $d_A=N_c^2-1$ is the dimensionality of the
adjoint representation and $\rho_i$ is the density of the scatterers.
Using $\rho_i=\rho_i^{\rm iso}/\sqrt{1+\xi}$
we obtain
\begin{equation}
\frac{1}{\lambda} = \frac{18\alpha_s T\zeta(3)}{\pi^2\sqrt{1+\xi}} \frac{1}{R(\xi)}
\frac{1+N_F/6}{1+N_F/4}
\end{equation}
where $N_F$ is the number of flavors.
For $\xi \rightarrow 0$ Eq.(19) reduces to well-known 
results~\cite{magdalenaprc}
\begin{equation}
\frac{1}{\lambda} = \frac{18\alpha_s T\zeta(3)}{\pi^2}
\frac{1+N_F/6}{1+N_F/4}
\end{equation}

\vskip 0.5cm
\begin{figure}
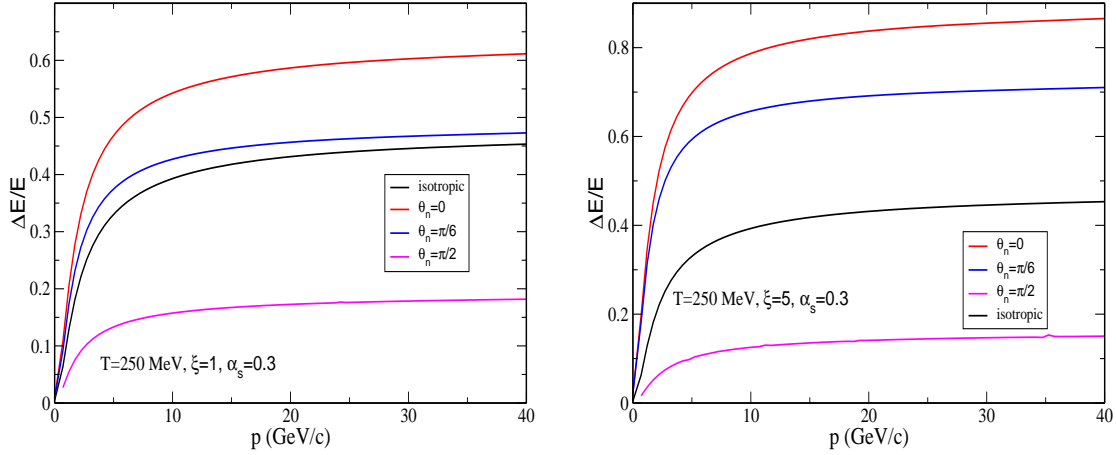

\begin{center}
\epsfig{file=dele_new_xi1_al.eps,width=7cm,height=6cm,angle=0}~~~~
\epsfig{file=dele_new_xi5_al.eps,width=7cm,height=6cm,angle=0}
\end{center}
\caption{(Color online) 
Same as Fig.~\protect\ref{fig1} for $\xi=1$ (left panel) and for $\xi=5$ 
(right panel). }
\label{fig2}
\end{figure}

It is evident that the changes from the isotropic media appear here
as the co-efficient ($R(\xi$)) of the Debye mass and the coefficient
($1/\sqrt{1+\xi}$) of the number density. In the limit 
$\xi\,\rightarrow\,0$ we recover all the previously known results
as may be checked from Ref.~\cite{magdalenaprc}.

\section{Results}

For the quantitative estimates of the fractional energy loss
in an anisotropic media, first we consider a plasma at a temperature
$T$ = 250 MeV with effective number of degrees of freedom
$N_F$=2.5 with the strong coupling constant $\alpha_s$=0.3 and $L$=5 fm.
We also note that the mean free path of the propagating parton
depends on the anisotropy parameter $\xi$ (see Eq.(19)).
The fractional energy loss for non-zero $\xi $ ($\xi$=0.5) 
for light flavour is shown in Fig.~(\ref{fig1}). 
As is evident from Eq.(16), the energy loss in anistropic media
depends on the angle of propagation of the fast partons
with respect to the anisotropy axis ($\vec n$). This is also illustrated in
Fig.~(\ref{fig1}).
It is observed that for non-zero value of 
the anisotropy parameter ($\xi$), the fractional energy loss
increases in the direction parallel to the anisotropy axis. 
However, away from the anisotropy axis, the fractional energy loss
decreases as the quark-quark potential is stronger in the anisotropy direction.

For higher value of the anisotropy parameter $\xi$ the results are shown
in Fig.~(\ref{fig2}). It is seen that the fractional energy loss increases
with $\xi$ in the anisotropy direction. For $\xi=1$ and $\xi=0.5$, 
the fractional energy 
loss increases marginally for $\theta_n=\pi/6$ and
it becomes larger for $\xi=5$ for the same value of $\theta_n$. 
However,
in the perpendicular direction the fractional energy loss decreases
substantially. {\bf It is to be noted that for small anisotropy the results
are almost similar to the case when the mean free path is independent of the
anisotropy parameter. However, for larger values of $\xi$ the result changes
reasonably as can be verified by calculating $\lambda$ from Eq.(19) for
larger anisotropy (see Fig.~(\ref{fig1})).}    

\begin{figure}
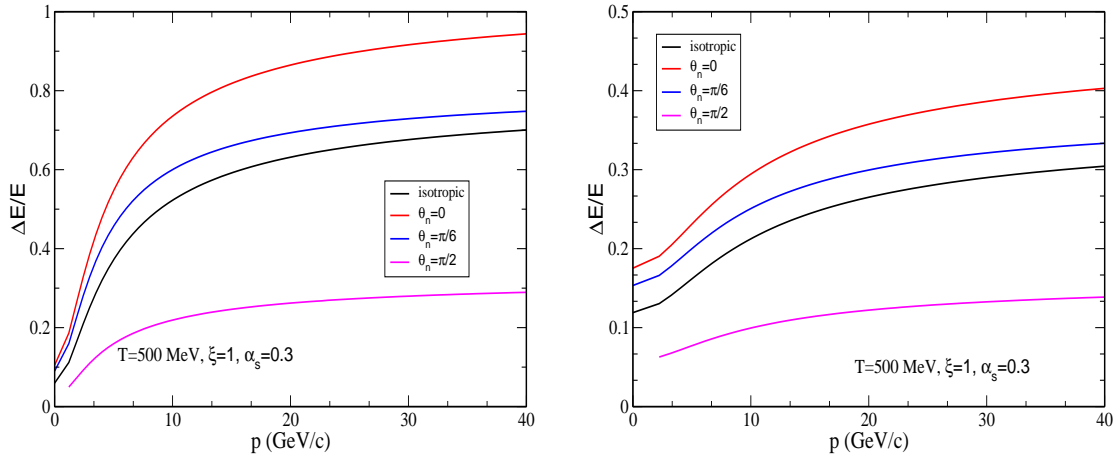

\begin{center}
\epsfig{file=dele_new_xi1_T500_cq_al.eps,width=7cm,height=6cm,angle=0}~~~~
\epsfig{file=dele_new_xi1_T500_bq_al.eps,width=7cm,height=6cm,angle=0}
\end{center}
\caption{(Color online) 
Same as Fig.~\protect\ref{fig1} for charm (left panel) and bottom quarks 
(right panel) with $\xi=1$ and $T =$ 500 MeV. }
\label{fig4}
\end{figure}

For the heavy quarks, i.e. for charm and bottom,
the results are shown in Fig.~(\ref{fig4}) for $\xi=1$. 
Similar to light quarks, we
find enhancement in the anisotropy direction as well as for
$\theta_n = \pi/6$. However,
for $\theta_n = \pi/2$ the energy loss (fractional) decreases for the
reasons mentioned earlier.


\section{Summary}
In this work, we have calculated the fractional energy loss
due to gluon radiation in an infinite size anisotropic media treating 
the scatterer as providing 
a screened coulomb-like potential. We have seen that the 
potential gets modified in anisotropic media. It is observed that
the fractional energy loss depends on the direction of propagation
of the fast partons with respect to the anisotropy axis as well
as on the anisotropy parameter $\xi$.
An enhancement is seen in the direction parallel to the anisotropy
direction $\vec n$, where as in the transverse direction it reduces
due to weaker quark-quark interaction.
It is also observed that for higher values of $\xi$, 
the fractional energy loss increases
for a given direction with respect to the anisotropy axis.
We also note that due to the dependency of the mean free path on the
anisotropy parameter, the energy loss increases as $\xi$ increases.
 
We do not include the recoil of the scatterer in this work. However,
this condition can be relaxed by incorporating the recoil corrections
which plays an important role as shown in Ref.~\cite{magdalenaprc}. This
will be included in future publication. Furthermore,
the finite size effect to the radiative energy loss
in anisotropic media would also be interesting to study.

The present calculation can be extended
to include the effect of the growth of unstable modes to obtain results valid
in a more realistic scenario as mentioned in the introduction. 
Inclusion of such effects might modify the quantitative
estimate of  nuclear modification
factor at RHIC and LHC energies.


\end{document}